\documentclass[9pt,twocolumn,twoside]{opticajnl}
\journal{opticajournal} 

\setboolean{shortarticle}{false}

\usepackage{lineno}

\usepackage{graphicx}
\usepackage{siunitx}
\DeclareSIUnit[quantity-product = ]\percent{\char`\%}
\usepackage{color}

\newcommand{\bluee}{black}

\newcommand*{\figref}[2][]{%
	Fig.~\hyperref[{fig:#2}]{%
		\ref*{fig:#2}%
		\ifx\\#1\\%
		\else
		(#1)%
		\fi
	}%
}

\newcommand*{\figureref}[2][]{%
	Figure~\hyperref[{fig:#2}]{%
		\ref*{fig:#2}%
		\ifx\\#1\\%
		\else
		(#1)%
		\fi
	}%
}

\newcommand*{\figuresref}[2][]{%
	Figures~\hyperref[{fig:#2}]{%
		\ref*{fig:#2}%
		\ifx\\#1\\%
		\else
		(#1)%
		\fi
	}%
}

\title{Supercharged two-dimensional tweezer array with more than 1000 atomic qubits}

\author[1]{Lars~Pause}
\author[1]{Lukas~Sturm}
\author[1]{Marcel~Mittenb\"uhler}
\author[1,2]{Stephan~Amann}
\author[1]{Tilman~Preuschoff}
\author[1]{Dominik~Sch\"affner}
\author[1]{Malte~Schlosser}
\author[1,*]{Gerhard~Birkl}

\affil[1]{Technische Universit\"{a}t Darmstadt, Institut f\"{u}r Angewandte Physik, Schlossgartenstra\ss e 7, 64289 Darmstadt, Germany}
\affil[2]{Current address: Max-Planck-Institut f\"ur Quantenoptik, Hans-Kopfermann-Stra\ss e 1, 85748 Garching, Germany}

\affil[*]{apqpub@physik.tu-darmstadt.de}

\begin{abstract}
We report on the realization of a large-scale quantum-processing architecture surpassing the tier of 1000 atomic qubits. By tiling multiple microlens-generated tweezer arrays, each operated by an independent laser source, we can eliminate laser-power limitations in the number of allocatable qubits. Already with two separate arrays, we implement combined 2D configurations of 3000 qubit sites with a mean number of 1167(46) single-atom quantum systems. The transfer of atoms between the two arrays is achieved with high efficiency. Thus, supercharging one array designated as quantum processing unit with atoms from the secondary array significantly increases the number of qubits and the initial filling fraction. This drastically enlarges attainable qubit cluster sizes and success probabilities allowing us to demonstrate the defect-free assembly of clusters of up to 441 qubits with persistent stabilization at near-unity filling fraction over tens of detection cycles. 
The presented method substantiates neutral atom quantum information science by facilitating configurable geometries of highly scalable quantum registers with immediate application in Rydberg-state mediated quantum simulation, fault-tolerant universal quantum computation, quantum sensing, and quantum metrology. \textcolor{blue}{Published as: Optica vol. 11, p. 222 (2024)}
\end{abstract}

\setboolean{displaycopyright}{false} 

\begin{document}

\maketitle
\section{Introduction}
In ambitious current efforts, various platforms, e.g., utilizing quantum states of ions, superconducting qubits, or neutral atoms, are being pursued with the objective to establish large-scale devices for quantum information science \cite{Alexeev2021,Altman2021,Cheng2023,Defenu2023}.
Seminal work in laser cooling and trapping of atoms has paved the way for optical tweezer arrays \cite{Birkl2001,Dumke2002} to render themselves as highly attractive configurations of inherently identical individual-atom quantum systems with large system size \cite{Kaufman2021}. Present implementations of optical tweezers created by spatial light modulators reach 1000 sites \cite{Ebadi2021} and assemble target geometries of 324 qubits \cite{Schymik2022}.
The precise control of the atomic qubits embraces single-qubit operations and the coherent application of Rydberg-mediated interactions of selectable strength thus enabling quantum simulation of spin Hamiltonians and gate-based universal quantum computation \cite{Adams2019,Browaeys2020,Henriet2020,Morgado2021,Shi2022,Cong2022,Kim2023,Wintersperger2023}.\\
Scaling up the system size is key for future progress towards leveraging quantum advantage. In this context, architectures for tweezer arrays based on passive microstructured optical elements \cite{Birkl2001,Dumke2002,OhldeMello2019,Huft2022,Hsu2022,Huang2023} offer the advantage of handling high optical power while producing stable arrays of laser foci for atom trapping which makes them extremely suitable for large-scale applications. Microfabricated lens arrays (MLAs) facilitate the massively parallelized generation of optical tweezers when illuminated with a laser beam of sufficient size \cite{OhldeMello2019,Schlosser2023}. As the number of lenslets per square centimeter readily reaches \num{100000} and MLA wafers with areas of several \num{100} square centimeters can be produced, they have enormous potential in terms of scalability, only limited by the available laser power \cite{Zappe2012,Voelkel2012,Schlosser2023}.
\begin{figure*}[t]
	\centering 
	\includegraphics[width=\linewidth]{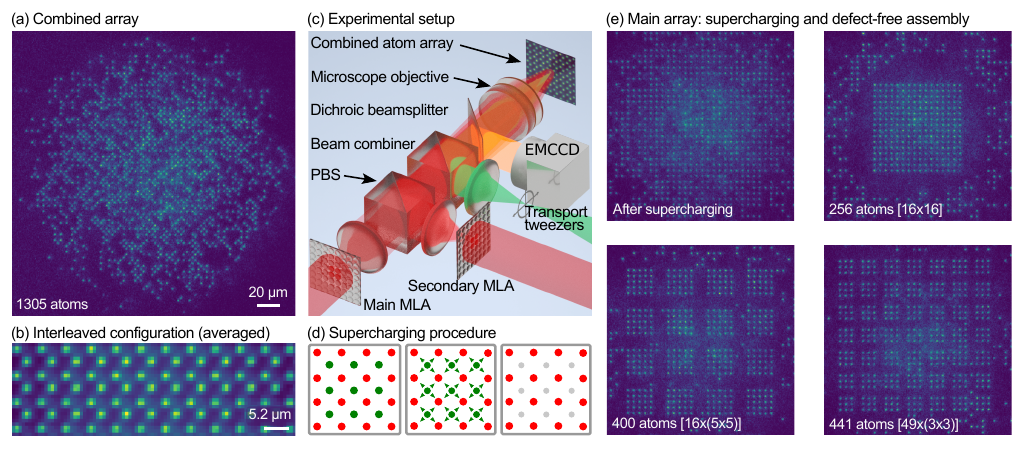} 
	\caption{%
		Large-scale registers of atomic qubits formed by interleaving two independent microlens-generated tweezer arrays.
		(a) In situ fluorescence image of a region that corresponds to 48$\times$48 sites of each original array showing 1305 single-atom $^{85}\text{Rb}$ qubits in the combined array of $\simeq 3000$ traps of sufficient depth.
		(b) Averaged fluorescence image depicting a central detail of the trap configuration for visualization of the interleaved configuration of two quadratic-grid arrays, each having a fundamental pitch of \SI{5.2(1)}{\micro\meter}.
		(c) Schematic experimental setup. The light fields of two microlens arrays (MLAs) are overlapped by a polarizing beamsplitter (PBS) and reimaged to form the combined atom array. Transport tweezers are used for qubit transfer from the secondary array to the main array and defect-free assembly of large-scale target patterns (see text for details).  
		(d) Pictogram of the supercharging procedure. Atoms from the secondary array (green dots) are used to fill defects in one of the four adjacent sites of the main array (red dots).
		(e) Section of 32$\times$32 sites of the main array. Directly after supercharging, the main array shows increased filling with \num{692} atoms. Defect-free assembly scores target patterns with up to \num{441} qubits.
		\label{fig:f1}}
\end{figure*}\\
In this article, we demonstrate a viable architecture eliminating this limitation and present two-dimensional (2D) arrays of approximately \num{3000} traps of sufficient depth. 
This unprecedented large-scale realization is achieved by the application of two separate quadratic-grid tweezer arrays that are created in parallel from independent laser sources and overlapped with high efficiency in an interleaved configuration. 
\figureref[a]{f1} displays an in situ fluorescence image of the resulting atom pattern showing \num{1305} single-atom qubits. A detail of the periodic atom arrangement is shown in \figref[b]{f1} as averaged fluorescence image.\\
In most optical tweezer setups, a single trapping-laser source is used (see Refs.~\cite{Graham2022,Singh2022a} for noteworthy exceptions). For a given trap size and laser detuning, the trap depth is proportional to the laser power per trap. Thus, the number of sites with sufficient depth is proportional to the available laser power.
However, for a single laser source this quantity is limited by various criteria. Starting with the maximum output power attainable with state-of-the-art laser technology at the required wavelength, additional limitations imposed by other components have to be considered:~The transmission efficiency of optical fibers is limited by stimulated Brillouin scattering and further restrictions are imposed by the damage thresholds as well as thermal limits of various active optical components, such as spatial light modulators used in many tweezer setups.
In the extended microlens-based platform of this work, the first two criteria, i.e., laser output power and fiber transmission efficiency, present the predominant limitations. Identifying laser sources operating in the wavelength range around \SI{800}{\nano\meter} (e.g., TiSa lasers) as a preferable choice for large-scale trapping arrays for rubidium atoms, the maximum achievable power at the output of a single-mode optical fiber available for creating tweezer arrays is currently on the order of several Watts. Considering typical losses in the optical components up to the plane of atom trapping, this results in about \num{1500} trap sites of sufficient depth in our setup. 
The limitation is lifted by implementing multiple tweezer arrays operated in parallel with all arrays being generated fully independently by separate laser sources and optical setups for the creation of the focal structures. Combining these in an interleaved, stacked, or tiled configuration, extends the achievable number of atomic qubits far beyond the limit imposed by one laser source. As shown in this article for two arrays, interleaving the main trapping array with a secondary array created in parallel from a second laser source of comparable output power and wavelength, allows for a doubling of the number of traps and a significant increase in accessible size of the qubit target pattern.
\section{EXPERIMENTAL SETUP}
A schematic of the experimental setup of our MLA-based platform is shown in \figref[c]{f1} (see Refs.~\cite{Dumke2002,OhldeMello2019,Schlosser2023} for additional details). Illuminating an MLA of $\SI{75}{\micro\meter}$ pitch and $\SI{1.1}{\milli\meter}$ focal length with a Gaussian laser beam generates a 2D array of laser spots in the focal plane of the MLA.
To relay the focal structure into the vacuum chamber and demagnify it, the focal plane is reimaged by an optical system consisting of a long-focal-length achromatic lens doublet and a microscope objective. This creates a quadratic-grid tweezer array with \SI{5.2(1)}{\micro\meter} pitch and \SI{1.0(1)}{\micro\meter} waist ($1/e^2$~intensity radius) which is loaded with laser-cooled atoms from a magneto-optical trap.
At a wavelength of \SI{799.5}{\nano\meter}, a laser power of \SI{0.5}{\milli\watt}, as typical for one of the central traps, results in a trap depth of $U=k_B\times\SI{0.5}{\milli\kelvin}$.
For all traps, the collisional blockade effect (\cite{Brown2019,Schymik2022}, and references therein) induces a maximum occupation of one laser-cooled $^{85}$Rb atom per site, but also about \SI{40}{\percent} of empty traps. Atoms in Talbot planes \cite{Schlosser2023} are removed by a resonant blow-away laser beam to a high degree.\\
For generating a multi-MLA tweezer array, in the current implementation we combine the light fields of two identical MLAs via a polarizing beamsplitter cube (PBS) in the collimated-beam section between achromatic lens and microscope objective.
A non-polarizing beam combiner is used for injecting steerable tweezers for atom transport and target-structure assembly.
\textcolor{\bluee}{Transport operations are carried out sequentially, i.e., atom-by-atom. Each move is performed by applying two intensity ramps of 200 µs duration each for atom extraction and delivery and an average transport velocity of \SI{16}{\micro\meter/\milli\second}.}
For atom detection, the fluorescence light of the trapped atoms is deflected to a highly-sensitive electron-multiplying charge-coupled device (EMCCD) camera by a dichroic beamsplitter, giving an average detection fidelity of \SI{99.8(2)}{\percent}.
\section{RESULTS}
One of the central achievements reported in this work is the demonstration of the scaling of the number of trapped individual atoms with the number of independent MLA-based arrays when operated in parallel in multi-array configurations. For the interleaved configuration of two arrays as detailed in \figref[b,d]{f1}, up to \num{1305} individual atoms have been recorded in the combined array (\figref[a]{f1}). The data set of \figref[]{f2} summarizes measurements of loading efficiency and atom number for the two arrays operated separately and in parallel. \figureref[a]{f2} displays the site-resolved loading efficiency of the two trap arrays in the reimaged focal plane (i.e., atom plane) when operated in parallel. We observe a circular plateau of efficient loading with a radius of roughly \num{22} pitches and an average loading efficiency of approximately \SI{40}{\percent}.
The extent of this plateau is related to the $1/e^2$ radius of \num{32} trap pitches of the Gaussian intensity envelope. The trap depth is reduced to a factor of $1/e$ relative to the central trap at a radial distance of \num{22} pitches. Beyond this distance, the traps become too shallow to hold individual atoms and the spread of light shifts from inner to outer traps starts to impede the global optimization of laser cooling parameters for atom loading and detection. For the combined array, this results in $\simeq 3000$ traps of sufficient depth.
\begin{figure}[t]
	\centering 
	\includegraphics[width=\linewidth]{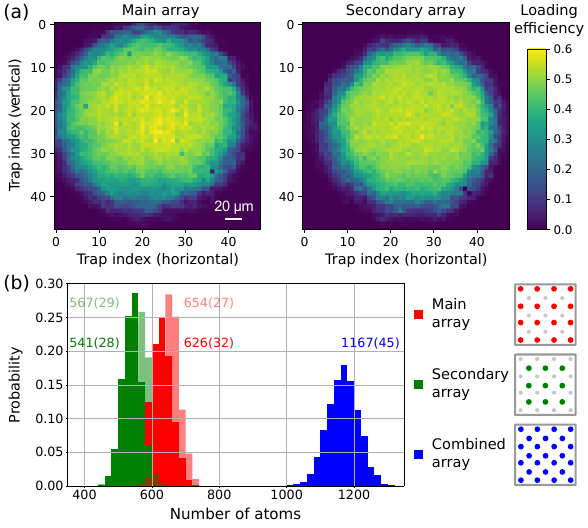} 
	\caption{
		(a) Site-resolved efficiency of single-atom loading of the two trap arrays when operated in parallel. Each pixel gives the color-coded efficiency value at the respective trap position within an 48$\times$48 trap array. The circular plateaus result from the requirements of sufficient trap depth for atom loading and uniformity in the occurring light shifts for atom detection.
		(b) Atom number statistics for separately operated (light color) and combined (dark color) tweezer arrays.
		\label{fig:f2}}
\end{figure}\\
In \figref[b]{f2} the number statistics of trapped atoms is presented.
In the combined setting with main and secondary array operated in parallel (blue), \num{1167(45)} individual atoms are recorded on average. In every single realization out of a total of \num{2000} repetitions, more than 1000 individual atoms have been detected. This number follows from \num{626(32)} atoms in the main array (dark red) and \num{541(28)} atoms in the secondary array (dark green). The corresponding numbers when loading atoms only into one of the two arrays, are shown in light colors, evaluating to \num{654(27)} atoms (main array) and \num{567(29)} atoms (secondary array), respectively. Thus, the presence of the respective other trap array reduces the number of trapped atoms by less than \SI{5}{\percent}.\\
The second central result of this work is the demonstration of the superior scaling of defect-free target clusters accessible with multi-MLA configurations. For this purpose, the main array of \figref[]{f2} is considered as quantum processing unit (QPU) with efficient scalability in the number of qubit sites and properties that qualify for the implementation of Rydberg-state mediated quantum information processing \cite{Schlosser2020}. The QPU is complemented by one or more secondary arrays serving as independently operated reservoirs, supercharging the QPU with additional single-atom qubits before or during operation. 
This architecture allows for scaling of the integral number of qubits almost proportional to the total number of operated arrays as presented in \figref[]{f2}. As a result, we overcome the limitations associated with most previous implementations where qubit registers are assembled typically by transport of atoms within a single array of optical tweezers.\\
The pictograms of \figref[d]{f1} illustrate the supercharging procedure for the interleaved configuration of \figref[a,b]{f1}: After initial loading of the combined array, the occupation of each site in both arrays is determined. Each site of the secondary array (green dots) serves as an on-demand reservoir for the neighboring sites of the main array (red dots). Following one cycle of insertion of reservoir atoms into unoccupied sites of a predefinable region of the main array using the transport tweezers, the secondary array is switched off.
In this fashion, the supercharged main array is created.
\textcolor{\bluee}{We use a heuristic algorithm that processes defects in the main array in sequence according to the trap index (see \figref[a]{f2}). After defining trivial moves for the cases where pairs of defects and reservoir atoms can be assigned without ambiguity, remaining defects are filled, if possible, without further prioritization, using the first indexed neighboring reservoir atom. In a Monte Carlo simulation of this procedure, the supercharged array reaches \SI{89}{\percent} filling, assuming \SI{50}{\percent} initial filling in the combined array and perfect insertion efficiency.\\}
\figureref[e]{f1} (after supercharging) shows a 32$\times$32 site section of the main array, which is the operation range of the transport tweezers, thus defining a QPU with \num{1024} qubit sites, with supercharging having been applied to the central 26$\times$26 site region, as typical for this work. The average insertion efficiency from secondary to main array is \SI{78}{\percent}.
Next, we implement up to \num{50} cycles of target-pattern assembly with an average duration of \SI{126}{\milli\second} in direct succession compensating for atom loss induced by the finite residence time of the atoms of \SI{10}{\second} and for the finite intra-array transport efficiency of \SI{89}{\percent}.
Empowered by this method, the subsequent successful defect-free assembly of various large-scale target patterns of up to 441 qubits is presented in the remaining panels of \figref[e]{f1}.
In \figref[]{f3}, we illustrate the advantage of the supercharging procedure. \figureref[a]{f3} visualizes the evolutionary assembly process and stabilization of a compact target cluster of 15$\times$15 sites (\num{225} atoms). Target assembly with the main array only and supercharged assembly are compared by depicting the respective cumulative defect-free success probabilities and filling fractions as a function of the number of executed assembly cycles.
Without supercharging, the cumulative success probability levels at a value of \SI{24}{\percent} after about \num{40} assembly cycles. With supercharging, the cumulative success probability increases to a value of \SI{35}{\percent}, still not exhibiting saturation at the maximum implemented number of \num{50} assembly cycles.\\
From similar measurements, we determined the success probabilities of even larger defect-free target clusters such as the patterns depicted in \figref[e]{f1}. 
For the \num{256} atom cluster in a compact configuration of 16$\times$16 sites, supercharging increases the cumulative success probability from \SI{8}{\percent} to \SI{14}{\percent}, whereas the \num{400} atom configuration of \num{16} clusters of 5$\times$5 sites and the pattern of \num{441} atoms in a configuration of \num{49} clusters of 3$\times$3 sites could be achieved for the supercharged configuration exclusively. For quantifying the improvement relative to our previous work \cite{OhldeMello2019}, we determined the cumulative success probability for a defect-free 10$\times$10 cluster. In this work we achieve a success probability of \SI{97}{\percent} as compared to \SI{3.1}{\percent} of Ref.~\cite{OhldeMello2019}.
\begin{figure}[t]
	\centering 
	\includegraphics[width=\linewidth]{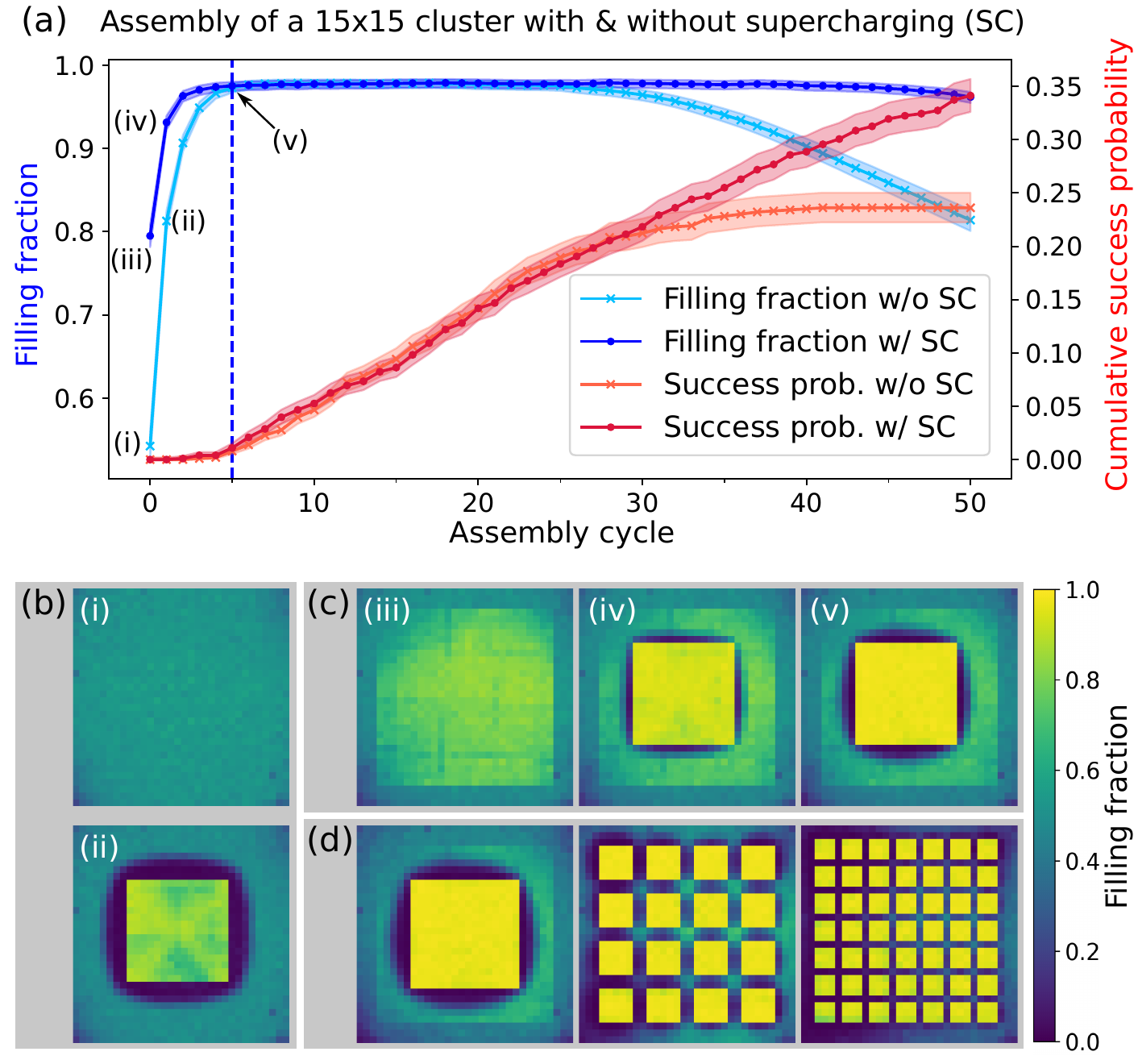} 
	\caption{
		(a) Cumulative success probabilities and filling fractions for a defect-free target pattern of \num{225} atoms in a cluster of 15$\times$15 sites. Supercharged assembly (w/ SC) with one cycle of atom insertion from the secondary array to the main array before the first assembly cycle is compared to assembly within the main array only (w/o SC).
		(b) Site-resolved filling fraction without supercharging before (i) and after (ii) the first assembly cycle.
		(c) Site-resolved filling fraction with supercharging before (iii) and after (iv) the first assembly cycle and after five assembly cycles (v).
		(d) Site-resolved filling fractions after five assembly cycles showing near-unity filling for all target patterns presented in \figref[e]{f1}. All depicted trap arrays comprise 32$\times$32 sites.
		\label{fig:f3}}
\end{figure}\\
\figureref[b,c]{f3} confirms the improvement achieved with supercharging by comparison of the site-resolved single-atom filling fractions of the main array before and after the first assembly cycle without (b) and with (c) prior insertion of atoms from the secondary array. The data points corresponding to the respective images (i)\,-\,(v) are labeled in \figref[a]{f3}.
The initial filling fraction for the supercharged section of 26$\times$26 sites is enhanced to \SI{74(2)}{\percent} as compared to \SI{53(2)}{\percent} obtained for loading the main array only. For the central 15$\times$15 target pattern, supercharging increases the initial filling fraction from \SI{54(3)}{\percent} (i) to \SI{80(3)}{\percent} (iii). A comparison of the site-resolved filling fractions after the first assembly cycle [(iv) in comparison to (ii)]
confirms the higher efficiency and reveals a significantly reduced depletion of atoms just outside the borders of the target structure for the supercharged configuration.\\
A plateau of almost constant filling fraction of \SI{98(1)}{\percent} in the 15$\times$15 target cluster is entered after five assembly cycles [vertical dashed blue line in \figref[a]{f3}] for both cases which for supercharging lasts until cycle 40 and without supercharging starts to decrease after cycle \num{25}. For supercharging, the site-resolved filling fraction after five cycles is shown in (v). The images in \figref[d]{f3} display the site-resolved filling fractions for the target structures presented in \figref[e]{f1} after five assembly cycles. The respective values of the filling fraction are \SI{97(1)}{\percent} [16$\times$16], \SI{98(1)}{\percent} [16$\times$(5$\times$5)], and \SI{95(1)}{\percent} [49$\times$(3$\times$3)].
For the compact clusters (15$\times$15 and 16$\times$16), the enhanced qubit reservoir outside the target pattern is still prominent in the images, resulting in a significantly prolonged stabilization phase with the above discussed plateau in the recorded filling fraction.
\section{DISCUSSION AND CONCLUSION}
In this work, we have implemented an experimental platform that demonstrates the enormous potential of MLA-based multi-array architectures for scaling up tweezer arrays of atomic qubits for quantum information science.
By interleaving two independently generated arrays, we achieved a number of more than \num{3000} trap sites in one plane, loaded with a mean number of \num{1167} individual atoms, and assembled in defect-free clusters of up to \num{441} atomic qubits.\\
With this distinctive micro-optics-based technology, tweezer arrays are expandable into extensive architectures of interleaved, stacked, and tiled multi-array configurations in a modular fashion:
Using multiple narrow-bandwidth dichroic beam combiners instead of a PBS, the focal planes of several MLAs, illuminated by light from different laser sources, can be superposed without losses, thus multiplying the number of qubit sites in interleaved configurations. 
Tiling a common focal plane with spot arrays created by several laterally displaced MLAs extends the size of the combined array beyond the size limit using a single array.
Even when utilizing only a single MLA, multiple trap arrays are created by separate laser beams that illuminate the MLA under slightly different angles of incidence, as shown in Refs.~\cite{Dumke2002,Schlosser2011,Schlosser2023}.
\textcolor{\bluee}{The geometric overlay of on the order of $10 \times 10$ beams on a single MLA utilizing angular displacement and the Talbot effect \cite{Schlosser2023} is feasible. 
In addition, using state-of-the-art optical components, the chromatic combination of 10 trapping laser fields with reasonable detuning is achievable.
Both upgrades will result in sufficient traps
to fill an accessible field of view of \SI{1}{\milli\meter\squared} size with more than \num{100000} tweezers.}
%
Displacing the focal planes of different MLAs axially results in stacked multilayer configurations with a predefinable distance between planes.
In three-dimensional architectures, scalability is further enhanced for a Talbot tweezer lattice~\cite{Schlosser2023}, giving stacked multilayer configurations with beneficial laser power scaling.\\
The presented architecture facilitates the inclusion of independent laser sources for 
generating multiple trap arrays which are combined in a scalable fashion with high efficiency. This vanquishes previous limitations by converting a hard technological limit into a linear dependence of the number of available qubits on the number of laser sources.
Due to the importance of scalability, we foresee broad impact on quantum information science with optically trapped atoms and molecules \cite{Kaufman2021}. In this context, it is important to note that our method is species-comprehensive, can be applied for various laser wavelengths in parallel \cite{Young2020,Singh2022a,Zhang2022} while retaining control of polarization \cite{Xia2015,Guo2020} and phase \cite{Kruse2010,Schlosser2023}, and is suitable for approaches utilizing incoherent trapping-light sources \cite{Huft2022}.\\
Future extensions include the adaption of techniques for enhanced initial preparation of individual atoms \cite{Brown2019,Aliyu2021,Schymik2022,Jenkins2022,Shaw2023} as well as the integration of continuously operated reservoir arrays for repeated resupply of atomic qubits \cite{Wang2020,Pause2023,Tao2023}.
\textcolor{\bluee}{In this work, the size of the assembled patterns is bounded by the finite residence time and the sequential nature of the atom transport with non-ideal efficiency.}
Technical revisions that expedite the assembly process, such as the implementation of parallelized atom transport \cite{Lengwenus2010,Bluvstein2022,Tian2023}, complemented by improved vacuum conditions, as ultimately achieved in cryogenic setups \cite{Schymik2022}, bear the potential for further propelling atomic quantum arrays to
\textcolor{\bluee}{qubit numbers of \num{e5}}
by exploiting the straightforward linear scaling with the number of laser sources, as introduced here.

\begin{backmatter}
\bmsection{Funding} Federal Ministry of Education and Research (BMBF) [Grant 13N15981], Deutsche Forschungsgemeinschaft (DFG -- German Research Foundation) [Grant No. BI 647/6-1 and BI 647/6-2, Priority Program SPP 1929 (GiRyd)], Open Access Publishing Fund of Technische Universit\"at Darmstadt.

\bmsection{Acknowledgments} We thank the labscript suite \cite{Starkey2013} community for support in implementing state-of-the-art control software for our experiments.

\bmsection{Disclosures} The authors declare no conflicts of interest.

\bmsection{Data Availability Statement} Data underlying the results presented in this paper may be
obtained from the authors upon reasonable request.

\end{backmatter}


\bibliography{1000_Atoms.bib}


\end{document}